\documentclass[12pt,preprint]{aastex}

\newcommand{\etal}{et~al.}
\newcommand{\mic}{~$\mu$m}

\newcommand{\methane}{CH$_4$}
\newcommand{\acetylene}{C$_2$H$_2$}

\newcommand{\wvn}{\,cm$^{-1}$}
\newcommand{\pcmsq}{$\,$cm$^{-2}$}

\begin{document}
\title{Warm HCN, \acetylene, and CO in the disk of GV Tau}
\author{E. L. Gibb, K. A. Van Brunt}
\affil{Dept. of Physics \& Astronomy, University of Missouri - St. Louis, 8001 Natural Bridge Rd, St. Louis, MO, 63121}
\author{S. D. Brittain}
\affil{Dept. of Physics \& Astronomy, Clemson University, Clemson, SC, 29631}
\author{T. W. Rettig}
\affil{Dept. of Physics, University of Notre Dame, Notre Dame, IN, 46556}

\begin{abstract}

  We present the first high-resolution, ground-based observations of
  HCN and \acetylene\ toward the T Tauri binary star system GV Tau.
  We detected strong absorption due to HCN $\nu_3$ and weak
  \acetylene\ ($\nu_3$ and $\nu_2 + (\nu_4 + \nu_5)^0_+$) absorption
  toward the primary (GV Tau S) but not the infrared companion.  We
  also report CO column densities and rotational temperatures, and
  present abundances relative to CO of HCN/CO $\sim$~0.6\% and
  \acetylene/CO $\sim$~1.2\% and an upper limit for \methane/CO $<$
  0.37\% toward GV Tau S.  Neither HCN nor \acetylene\ were detected
  toward the infrared companion and results suggest that abundances
  may differ between the two sources.

\end{abstract}


\section{Introduction}

GV Tau (Haro 6-10, IRAS 04263$+$2426) is an unusual young T Tauri
binary system partly embedded in the L1524 molecular cloud.  It is one
of a small number of young binaries for which the primary (GV Tau S)
is optically visible and the infrared companion (IRC, GV Tau N),
located 1$\farcs$2 to the north, is strongly extincted.  The spectral
energy distribution (SED) of the primary is flat or rising in the
1--100~\mic\ range, suggesting that it is a class I object
\citep{kore99}.  \citet{lein89} found that the IRC was generally
brighter than the primary at wavelengths longer than $\sim$4~\mic. It
has also been found that the system is variable, particularly in the
near infrared, and on timescales as short as a month
\citep{lein01,kore99}.  \citet{lein01} found that the primary became
redder as it became fainter at K.  Also, the depth of the 3~\mic\ ice
band was found to vary with time, and their results suggested that an
increase in optical depth corresponded to a decrease in K brightness.
Both of these measurements are consistent with the dominant mechanism
of the variability of GV Tau S being due to inhomogeneities in
circumstellar material that result in changes in extinction.  The
behavior for the IRC is more complex and suggests the possibility of
variable accretion mechanisms \citep{lein01}.

The system shows a parsec-scale Herbig-Haro flow with a well-defined
axis, as well as other smaller flows, one of which is associated with
the IRC, and one that does not appear to be associated with either
object \citep{devi99}.  Near-infrared emission images suggest the
presence of a flattened circumbinary envelope with a semi-major axis
of $\sim$1000--1500~AU seen nearly edge-on as the source of the
observed polarization and part of the extinction toward both objects
\citep{mena93}.  We note that their observations do not exclude the
additional presence of circumstellar disks around one or both objects.

In this paper we report high-resolution near-infrared detections of
CO, HCN, and \acetylene\ toward GV Tau S.  For the IRC, CO was
detected in absorption, but HCN and \acetylene\ were not. In section 2
we discuss the observations and data analysis.  In section 3 we
present the results of analysis for HCN, \acetylene, and CO, along
with upper limits for \methane.  Results are discussed in section 4.

\section{Observations and data reduction}

High-dispersion, infrared observations of GV Tau were taken with the
cryogenic echelle spectrometer NIRSPEC at the 10-meter W. M. Keck
Observatory on Mauna Kea, Hawaii \citep{mcle98}.  NIRSPEC provides a
resolving power of $\sim$25,000, which allows for the clean separation
of fundamental and excited states of $^{12}$CO, $^{13}$CO, and
C$^{18}$O as well as the $\nu_3$ band of HCN and two strong bands of
\acetylene ($\nu_3$ and $\nu_2 + (\nu_4+\nu_5)^0_+$).
Table~\ref{table1} presents the spectral settings, wavelength
coverage, and integration times for the spectral orders analyzed in
this paper.  The M- and K-band data covering the fundamental and
overtone transitions of CO, respectively, were acquired in March and
August 2003.  KL band spectra were obtained on February 17-18, 2006 to
search for gas phase absorptions due to organic species toward GV Tau.
The infrared companion of GV Tau was observed simultaneously during
the February 17, 2006 and March 2003 observations, enabling us to
characterize the spectra toward both objects.

\begin{figure}
\epsscale{0.9} 
\plottwo{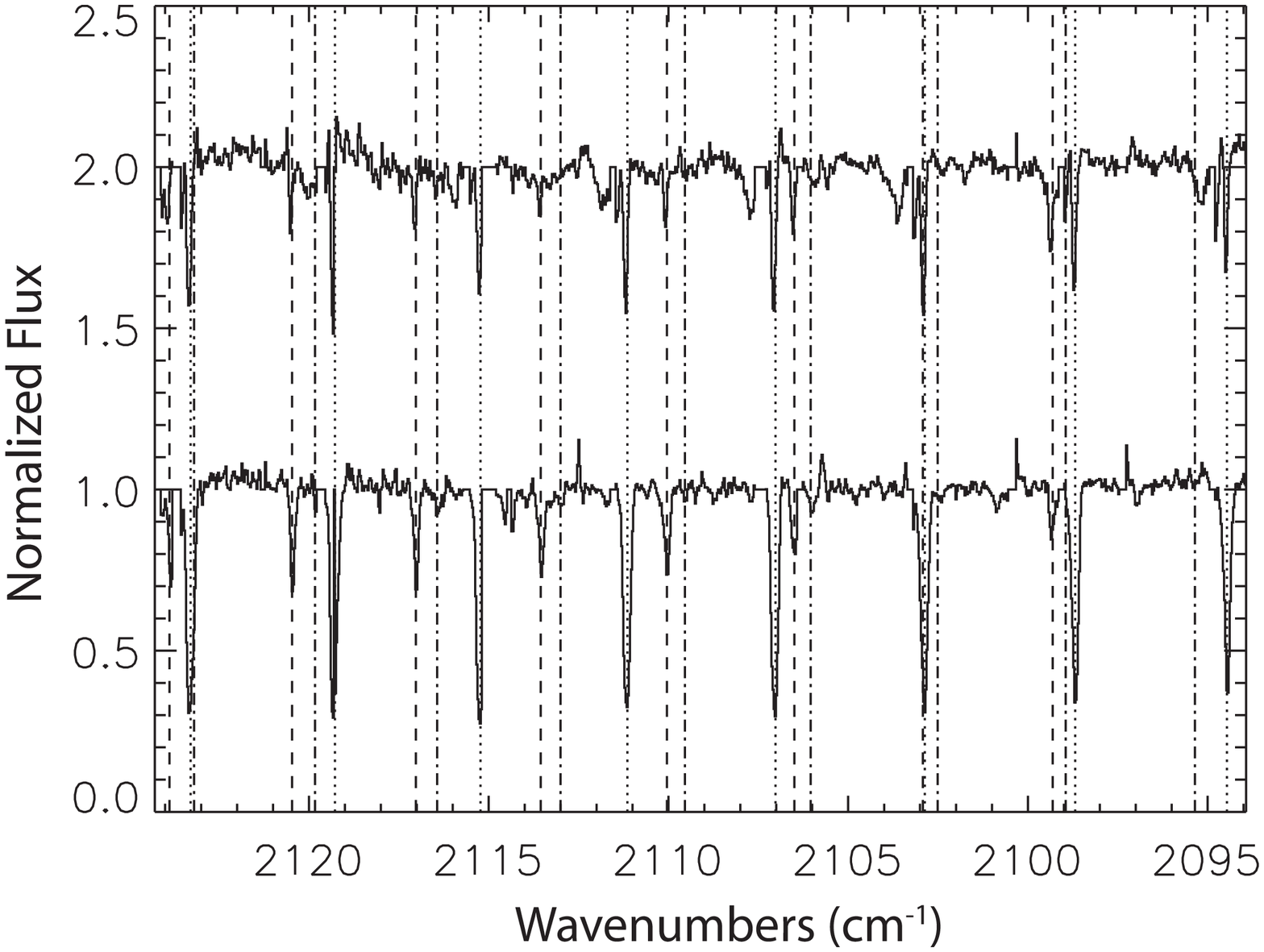}{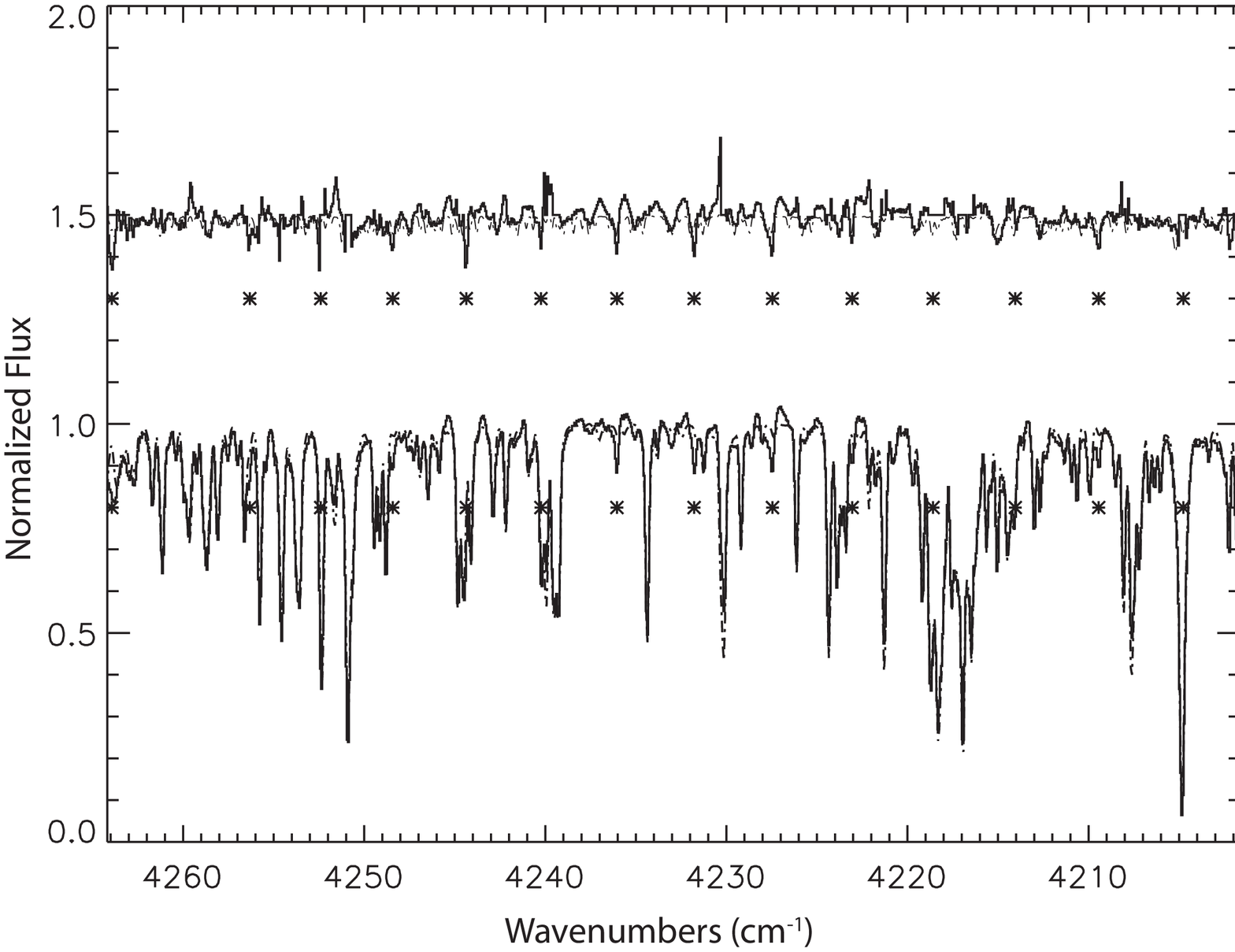}
\caption{(a) Sample normalized residual M-band spectrum showing CO
  absorption toward GV Tau S (bottom) and the IRC (top). The dotted
  lines indicate positions of the $^{12}$CO lines, the dashed lines
  indicate $^{13}$CO, and the dot-dashed lines indicate C$^{18}$O. (b)
  Normalized K band spectrum of GV Tau S with the telluric model
  overplotted (dot-dash line).  Above is the residual with the
  Arcturus spectrum overplotted, convolved to the resolving power of
  NIRSPEC, veiled and Doppler shifted to the geocentric velocity of GV
  Tau.  Asterisks indicate the positions of CO (2-0) absorption
  lines. \label{fig1}}
\end{figure}

\begin{figure}
\epsscale{1.0}
\plotone{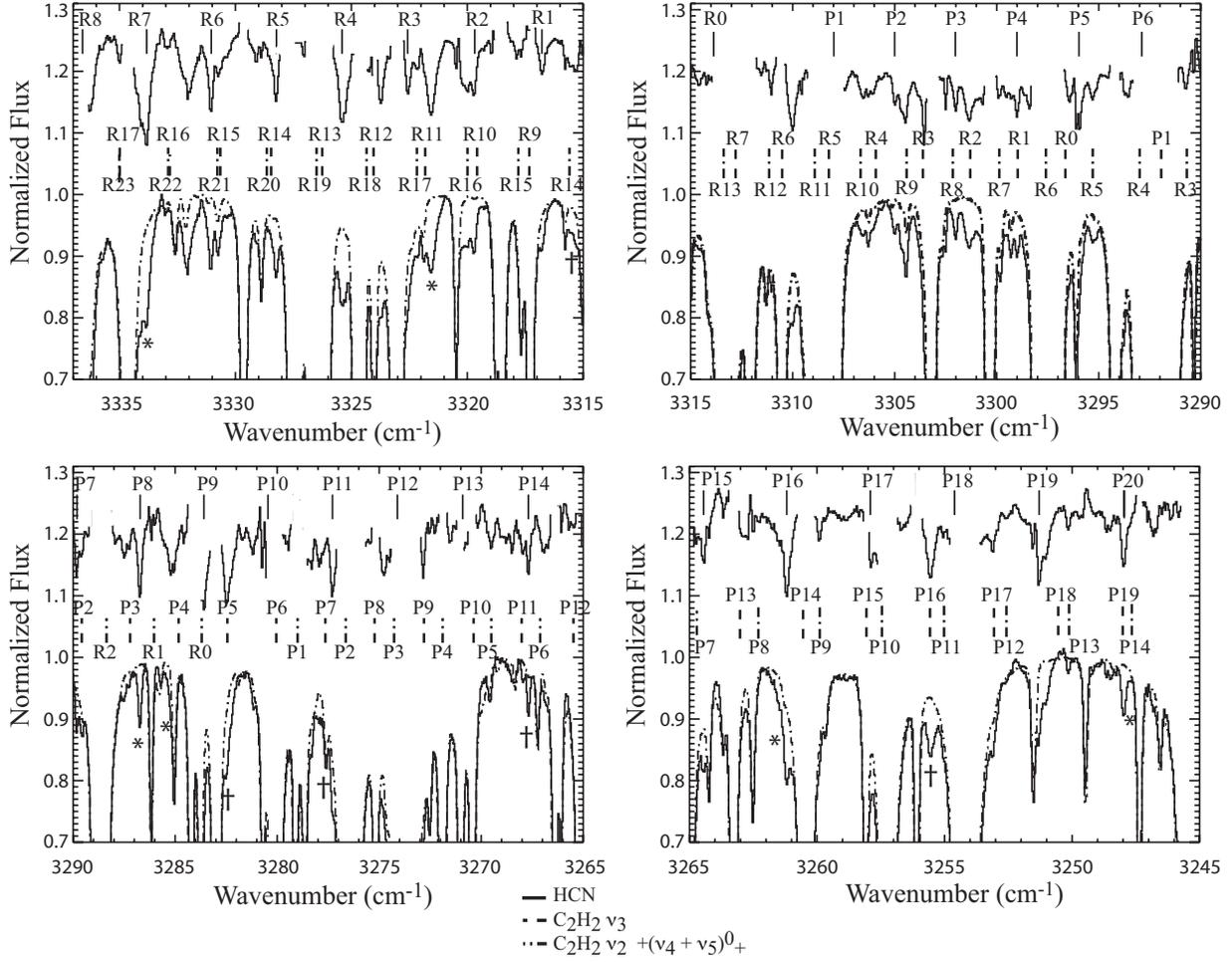}
\caption{GV Tau S, KL Order 25 spectrum showing the positions of HCN
  and \acetylene\ lines. Solid black ticks are the HCN $\nu_3$ band,
  dashed ticks indicate the \acetylene\ $\nu_3$ band, and dot-dashed
  ticks indicate the \acetylene\ $\nu_2$ + ($\nu_4$ + $\nu_5$)$^0_+$
  band. Asterisks and crosses denote Doppler shifted positions of
  stellar OH and NH, respectively. \label{fig2}}
\end{figure}

Our reduction and analysis procedures are discussed in
\citet{disa01,brit03}.  We summarize them below.  Observations at L
and particularly at M band are dominated by a thermal ($\sim$300 K)
background with sky emission lines superposed.  In order cancel
atmospheric and background effects, we nodded the telescope between
positions one quarter of the way from the top and bottom of the slit,
our "A" and "B" positions, respectively.  Each exposure corresponded
to one minute of integration time.  Telluric lines were cancelled to
first order by combining observations as \{A-B-B+A\}/2. A series of
flats and darks was taken for each grating setting.  After
flatfielding and dark subtraction, systematically hot and dead pixels
and cosmic ray hits were removed and the data were resampled to align
the spectral and spatial dimensions along rows and columns,
respectively. Examples of extracted spectra are shown in
Figures~\ref{fig1} and \ref{fig2}.  The individual lines of each
molecule are labeled.

The atmospheric transmittance function was modeled using the Spectrum
Synthesis Program \citep[SSP]{kund74}, which accesses the updated
HITRAN 2000 molecular database \citep{roth03}.  For each grating
setting, the column burdens of atmospheric species, the spectral
resolving power, and the wavelength calibrations were established by
fitting the optimized model.  To extract our spectral absorption
features, we subtracted the model, convolved to the resolution of the
instrument, from the spectral extract.  This results in a residual
that is still convolved to the telluric spectrum.  For intrinsically
narrow lines, we obtain the true line intensity incident at the top of
the terrestrial atmosphere by dividing each spectral line by the
transmittance at the Doppler shifted position using the fully resolved
model.  The optimized model is shown as the dashed line in
Figures~\ref{fig1} and \ref{fig2} and has been found to reproduce the
telluric spectrum accurately.

We note that the relative continuum brightnesses of the primary and
companion varied significantly between our 2003 and 2006 observations.
In particular, the IRC was fainter than GV Tau S through the
2--5~\mic\ region.  Both objects are known to vary in the infrared
\citep{lein01} and the variations are thought to be due primarily to
variable amounts of intervening gas and dust, essentially clumpy
circumbinary material. \citet{lein01} reported that GV Tau S got
slightly fainter at M band over the period from 1993-2000 while the
IRC was found to get brighter during the same time interval.  A
similar trend was seen for both objects at L'.  In SpeX data acquired
in November 2004, the infrared companion was substantially fainter
than GV Tau S throughout the 2-5~\mic\ region.  This was also true for
our 3.3~\mic\ observations in 2006.  Nonetheless, we were able to
extract spectra from both GV Tau S and its IRC.

The extinction toward GV Tau S is relatively small (A$_V$ $\sim$5.6
mag) and thus we may also be detecting stellar photospheric absorption
lines.  To check for possible photospheric lines, we compared our
spectral type K3 GV Tau spectra \citep{good86} to a high-resolution K2
spectrum of Arcturus \citep{hink95}.  The Arcturus spectrum was
Doppler shifted to the geocentric velocity of GV Tau as determined by
our HCN and CO observations, which were found to be consistent with
the heliocentric velocity measured by \citet{whit04}, and convolved to
the resolution of our NIRSPEC data.  Broad (consistent the rotational
broadening reported by \citet{whit04}) absorption lines corresponding
to photospheric OH and NH were present toward GV Tau S (indicated by
crosses and asterisks in Figure~2).  Contaminated lines were
eliminated from our analysis.  The remaining absorption features were
measured by fitting a Gaussian profile to calculate the equivalent
width.

\section{Molecular Absorptions Toward GV Tau}

\subsection{CO gas spectra}

\subsubsection{GV Tau S}
For GV Tau S, the fundamental ro-vibrational lines of $^{12}$CO,
$^{13}$CO, and C$^{18}$O near 4.67~\mic\ and the overtone bands near
2.3~\mic\ were observed in absorption in March and August 2003,
respectively (Figure 1).  Absorption line studies have the advantage
of measuring physical conditions in a pencil-beam column of gas along
the line of sight to the emitting source, usually the star and inner
accretion disk.  This allows us to sample a long path length through
the edge-on circumbinary envelope and circumstellar disk.  The
rotational temperatures and column densities of the different bands
and isotopes of CO were derived using a population analysis and are
presented in Figure~3 and Table~2.  The overall rotational temperature
found for CO toward GV Tau S is $\sim$200~K, somewhat warmer than that
found for HCN (section 3.2), though within 3-sigma for the overtone
lines and lines of C$^{18}$O, and consistent with the T$_{rot}$ found
for \acetylene\ (section 3.4).  This is also similar to the warm
($\sim$100~K) temperatures we find toward other flared disk systems
\citep{rett06} and implies that the gas is located in the inner,
potentially planet forming ($\sim$10~AU or so) region of the
circumstellar disk rather than in the more distant interstellar or
circumbinary material.  The column density of $^{12}$CO was found to
be (5.9 $\pm$ 1.2) x 10$^{18}$~\pcmsq\ based on an analysis of the
overtone lines, which are optically thin.  The fundamental lines of CO
are optically thick and require a curve of growth analysis.

We found $^{12}$CO/$^{13}$CO to be 54 $\pm$ 15, consistent with that
measured toward Orion A \citep[67 $\pm$ 3]{lang90} and that found
toward the young star HL Tau \citep[76 $\pm$ 9]{brit05}.  The
$^{12}$CO/C$^{18}$O ratio is 420 $\pm$ 170, somewhat lower than toward
Orion \citep[$\sim$660]{lang90}, HL Tau \citep[800 $\pm$ 200]{brit05},
and the canonical interstellar value of 560 $\pm$ 25 found by
\citet{wils94}. We would expect the isotopic abundance ratios to
reflect the isotopic composition of the dense cloud unless influenced
by a mechanism such as selective dissociation \citep{lyon05}, which
may occur since the isotopomers are not self-shielded as effectively
as $^{12}$CO.  This effect does not seem to be evident toward GV Tau.

\subsubsection{IRC}

The infrared companion was observed at M band in March 2003, but was
not placed in the slit during the K-band observations.  The IRC was
resolved and $^{13}$CO was detected, though C$^{18}$O was not.  We
estimate that the $^{12}$CO column density is
$\sim$3x10$^{18}$~\pcmsq, assuming that the $^{12}$CO/$^{13}$CO ratio
toward the two objects is the same.  This assumption seems reasonable,
given that the $^{12}$CO lines toward the IRC are about half as strong
as those toward GV Tau S, and in order to bring N($^{12}CO$) for the
IRC in agreement with the primary, the $^{12}$CO/$^{13}$CO ratio would
have to be a factor of two higher.  Future observations of the
overtone CO lines toward the IRC are planned to directly measure
$^{12}$CO.

\subsection{HCN gas}

\subsubsection{GV Tau S}

Rather strong absorption lines due to the HCN (100-000) band near
3.0~\mic\ were detected toward GV Tau S.  Spectral extracts are shown in
Figure 2.  The geocentric Doppler shifted positions (+49 km/s) of the
HCN absorption lines and the specific identifications are indicated by
solid ticks.  We performed a rotational analysis on absorption
features that were at transmittance $>$80\%, were not blended with
\acetylene\ transitions (also indicated in Figure 2), and were clear
of contamination from stellar absorption features as determined from a
comparison to the similar spectral type high-resolution Arcturus
spectrum of \citet{hink95}.  There were a total of 10 lines that
satisfied these criteria, ranging from J=0-7.

Einstein A's and line positions were taken from the ab initio line
list calculated by \citet{harr02}, which expanded on previous line
lists and improved accuracy by using the most accurate potential
energy surface and dipole moment surface information available.  The
population diagram is shown in Figure 4.  The Doppler shifted
positions and equivalent widths used are given in Table 3. 

We note that the lines in our analysis are optically thin for most
reasonable values of the intrinsic line widths (b).  For example, the
R6 line is optically thin ($\tau < 1$) for all b $>$ 1.7 km/s.  For
R1, $\tau > 1$ is achieved only for b $<$ 0.9 km/s.  \citet{brit05}
found b = 1.3$\pm$0.1 for CO toward HL Tau.  \citep{lahu06} find that
no good fit can be made to the \acetylene\ and CO profiles for b $<$ 2
km/s.  \citet{boog02} find that 0.8$<$b$<$1.5 km/s toward L1489 to
explain the CO observations.  If we make a reasonable assumption that
b$>$1 km/s, then our absorptions are optically thin.  This is also
supported by the linearity of the population diagram.  From this, we
derive a rotational temperature of 115$^{+11}/_{-10}$~K and a total
HCN column density of (3.7$\pm$0.3) x 10$^{16}$~\pcmsq\ toward GV Tau
S.  Comparing this to the CO column density (section 3.1) provides an
HCN/CO $\sim$ 0.63\%

\subsubsection{IRC}

Interestingly, HCN absorption was not seen toward the infrared
companion star.  If we assume the same T$_{rot}$ as found for the
primary (115~K), the 3-sigma upper limit for the HCN column density
toward the infrared companion is 4.8 x 10$^{15}$~\pcmsq, a factor of
about 8 lower than the primary.  The estimated CO column density
toward the IRC is a factor of two less than toward GV Tau S.  If we
assume that the CO absorption originates in the inner circumstellar
disk of each object, then the resulting difference in HCN column
densities implies that HCN is at least a factor of four less abundant
toward the IRC.  This result is important.  The fact that the HCN
column density is significantly lower toward the IRC is a strong
constraint that argues for the HCN being located in the warm inner
disk of the primary, rather than in the interstellar medium or the
circumbinary material that surrounds both objects.

This result may suggest that there are significant compositional
differences between the inner disk ($\sim$10~AU or so) material toward
the two objects.  However, we point out that caution must be used when
comparing the primary to the IRC. There is no a priori reason to
assume the same gas temperature toward both objects if the gas is
associated with a circumstellar disk rather than the circumbinary
material.  The IRC was not located in the slit during our KL2
observation (Table~1).  This means that only the J$<$8 transitions of
HCN were covered.  Due to the fact that only low J lines were sampled
toward the IRC, the upper limit is less constrained for higher
temperatures.  For example, if we assume T=400~K, then the upper limit
for N(HCN) becomes $\sim$1.1x10$^{17}$~\pcmsq.

\subsection{\methane\ Gas}

We attempted to detect \methane\ in our KL, order 25 setting, which
covered several transitions.  With a geocentric Doppler shift of +49
km/s during the observations, the lines should have been shifted out
of the telluric features to transmittances ranging from $\sim$60\% for
P2 up to $\sim$80\% for R0 and P1 to 86\% for R1.  From this, we
calculated a 3-sigma upper limit for the \methane\ column density of
$<$2.2 x 10$^{16}$~\pcmsq\ using the methodology discussed in
\citet{gibb04} and assuming the same rotational temperature as that
found for HCN (115~K).  When compared with the HCN and CO we find that
\methane/HCN $<$ 60\% and \methane/CO $<$ 0.37\%.  This is not as well
constrained as the gas phase \methane/CO ratio found for HL Tau
\citep{gibb04}, but it is still substantially smaller than the overall
\methane/CO ratio found for comets \citep{gibb03} and somewhat lower
than reported for \methane/CO (ice+gas) for massive YSOs by
\citet{boog97,boog98}.

\subsection{\acetylene\ Gas}

There are two comparably strong bands of acetylene in the 3~\mic\
region: the $\nu_3$ band centered at 3294.84~\wvn\ and the $\nu_2$ +
$(\nu_4 + \nu_5)^0_+$ band centered at 3281.90~\wvn\ (see
\citet{jacq03} and references therein).  Ordinarily, the $\nu_3$ band
would be 1100 times stronger than the combination mode. However, a
Fermi resonance between the two bands causes the combination band to
become slightly stronger than the $\nu_3$ band \citep{vand93}.  

From the analysis of the 3~\mic\ spectra (Figure 2), a number of
unblended lines of \acetylene\ are detected at relatively good
transmittance ($>$80\%).  The positions of individual \acetylene\
lines are indicated in Figure 3 (dot-dash ticks). We analyzed 15
unblended lines with high transmittance from the $\nu_2$ + $(\nu_4 +
\nu_5)^0_+$ band and 4 lines of the $\nu_3$ band to determine the
column density (Figure 5, Table 4).  We also checked for possible
contamination due to stellar photospheric features by comparison with
an Arcturus spectrum.  The rotational temperature analysis yielded a
T$_{rot}$ = 170$^{+19}/_{-16}$~K, which is consistent with those found
for HCN and CO within the uncertainty, and a total column density of
(7.3$^{+0.1}/_{-0.2}$) x 10$^{16}$~\pcmsq.  These values are
consistent with the non-detection of the $\nu_3$ P18 line, which is at
97\% transmittance and unblended.  The predicted equivalent width of
P18, based on the fit in Figure 5, is 0.0011~\wvn, within the 1-sigma
noise limit of 0.0014~\wvn.
 
\section{Discussion}

We investigated the composition of several organic molecules toward
the binary T Tauri system GV Tau.  Such species as HCN, \acetylene,
and \methane\ are key to understanding the chemical compositions and
evolution of the volatile material that becomes incorporated into the
planet forming regions of disks around young stars. We detected HCN
and \acetylene\ toward the primary and provide upper limits for the
IRC.  This is only the second reported detection of these species in
the gas phase toward a low mass T Tauri star. The first reported
detection of these molecules was via Spitzer IRS observations toward
IRS 46 in the $\rho$ Ophiuchus cloud \citep{lahu06}.  The warm
temperatures ($\sim$400~K for HCN and $\sim$800~K for \acetylene)
suggested a possible disk origin for the absorptions, but were not
definitive as \citet{lahu06} could not distinguish whether the gas was
located in the disk or a jet.  

\subsection{Location of the Gas Toward GV Tau}

In February 2006, HCN was detected toward the primary (GV Tau S,
Fig.~\ref{fig2}).  However for the IRC, which was observed
simultaneously, the lack of a detection suggests a significant lower
total column density if we assume the same rotational temperature.
Similarly, the CO spectra toward GV Tau S and the IRC are
significantly different, suggesting a factor of two difference in
column density, resulting in a factor of at least four lower abundance
toward the IRC.  This result is a strong indication that the HCN is
located close to GV Tau S, within the circumstellar flared disk, and
not associated with the IRC (and therefore not in the circumbinary
material).  The warm rotational temperatures of $\sim\!$100-200~K
determined for HCN, \acetylene, and CO also suggest a location in the
inner region of the disk that surrounds GV Tau S rather than in the
circumbinary envelope.  A jet origin can likely be eliminated as the
Doppler shifts of stellar photospheric absorption lines agree with
those found for the intervening molecular material to within $\sim$5
km/s.  We derive a heliocentric radial velocity of 21$\pm$5 km/s for CO
and 19.5$\pm$3.4 for HCN. This is similar the heliocentric Doppler
shift of 13.4$\pm$3.8 km/s reported by \citet{whit04} using high
resolution optical spectra of stellar emission lines.  The combination
of velocity information, significantly different compositions toward
closely spaced objects, and warm rotational temperatures combine to
argue that the HCN, \acetylene, and CO are located in the disk of GV
Tau S.

If we assume that the near-infrared variability toward GV Tau is due
primarily to cooler material in the outer disk or circumbinary
envelope rather than the warm material where the CO, HCN, and
\acetylene\ absorptions originate, as suggested by the 3~$\mu$m ice
feature variability, then we can compare our column densities of HCN
and \acetylene\ to CO (see Table~2).  We must be cautious with such an
interpretation, however, since our CO (2 and 5~\mic) observations and
our 3~\mic\ observations were taken at different times and the sources
are both highly variable in the infrared over short time scales.  This
could affect the amount of gaseous material in the line of sight.  For
example, tidal interactions between the two stars, which are only
separated by about 170 AU, may affect the gas structure close to the
stars.  Perhaps the temperature and density structures of the inner
disks differ, resulting in dissimilar chemical evolution.  It may also
be possible that an inner circumstellar disk of the IRC is more
inclined than the nearly edge-on GV Tau S and that we are seeing warm
CO in the flared disk.  The molecules in disk atmospheres are less
protected from energetic photons than those in the midplane, leading
to higher photodissociation rates.  Collisional dissociation in the
inner disk may act to remove hydrogen atoms from molecules like
\acetylene\ and HCN.  Also, the increased production of ions in disk
atmospheres and the resulting ion-molecule chemistry will further
modify the composition.  The extent to which these processes occur is
dependent on the local ionizing radiation field, temperature, and
density (see papers by \citet{will98,marc02,aika99,aika02} for
discussion of chemistry in protoplanetary disks).  Hence, the
chemistry is dependent on radial and vertical distances in the disk
and our measurements are correspondingly dependent on viewing
geometry.  Additional observations to simultaneously measure CO and
HCN toward GV Tau are planned to address this issue and to investigate
the possibility of variations in column densities of gaseous species.

\subsection{Compositional Comparisons}

It is interesting to compare our results for GV Tau S to those found
toward other young stars, models of chemistry in protoplanetary disks,
and comets in our own solar system. Assuming that CO, HCN, and
\acetylene\ originate in the same region, we found a \acetylene/HCN
ratio of $\sim$2.0 toward GV Tau S and HCN/CO $\sim$ 0.63\% and
\acetylene/CO $\sim$ 1.2\%.  \acetylene/HCN was found to be $\sim$0.6
toward IRS 46 \citep{lahu06}, another low mass object, assuming that
both species originate in the same region.  The 13-14~\mic\ Q-branches
and band heads of \acetylene\ and HCN were studied by ISO
\citep{lahu00} toward massive YSOs.  In general, \acetylene/HCN for
the warm component was found to be $\sim$40-50\% toward most
objects. Interestingly, HCN was found to be more abundant
($\sim$2.5\%) toward IRS 46 compared to CO (this is a lower limit).
This can be seen graphically in Figure~\ref{fig6}.  In that figure,
\acetylene/HCN is indicated by asterisks while the abundances of
\acetylene\ and HCN relative to CO (in \%) are indicated by filled
squares and diamonds, respectively.

\citet{marc02} modeled molecular distributions of abundant species in
the inner 10 AU of protostellar disks surrounding T Tauri stars.
\citet{gibb04} measured an upper limit for \methane/CO toward HL Tau
that was significantly lower than predicted by the model.  The same
was found for GV Tau S, where \methane/CO $<$ 0.0037 is much smaller
than the \methane/CO $\sim$1 predicted by Markwick through most of the
inner disk.  However, we find that HCN/CO, \acetylene/CO, and
\acetylene/HCN ratios toward GV Tau S are in general agreement with
the range of abundances found in \citet{marc02} (see
Figure~\ref{fig6}).  Our rotational temperatures ($\sim$100-200~K) are
also consistent with the midplane temperatures in the inner $\sim$5~AU
of the protoplanetary disk in the Markwick model.  We note, however
that we are likely sampling a region above the midplane, possibly in a
disk atmosphere, which may be heated by various mechanisms such as
viscous accretion, X-rays, or an X-wind (see \citet{glas04} for a
discussion of disk heating).  This would move the location of the gas
outward by an amount depending on the local environment.  Nonetheless,
the similarity in abundances between \citet{marc02}, GV Tau, and IRS
46 is interesting and should be pursued further.

Another interesting comparison is with comets in our own solar system.
Comets are generally considered to be the most pristine objects in the
solar system and are thought to have formed in the giant-planet region
(from $\sim$5-30~AU).  Infrared spectroscopy has routinely measured
abundances of CO, HCN, and \acetylene\ toward comets for the last
decade (see \citet{mumm03} and references therein).  HCN and
\acetylene\ abundances measured in comets (compared to water) are
fairly consistent over the Oort cloud population measured to date and
the \acetylene/HCN ratio is found to be typically $\sim$0.9-1.1,
reasonably consistent with the \acetylene/HCN ratios of 0.6 and 2.0
found for IRS 46 and GV Tau, respectively.  The CO abundance in comets
has been found to be highly variable, likely as a result of thermal
history and processing of comets in the protoplanetary disk.  The
resulting HCN/CO and \acetylene/CO ratios range from $\sim$0.01-0.16.
The HCN/CO and \acetylene/CO ratios found in GV Tau and IRS 46 are
consistent with those found for CO rich comets.  It is possible that
the more volatile CO was preferentially lost in comets for which the
HCN/CO or \acetylene/CO ratios are high.  If we assume that the comets
with the greatest abundance of CO more closely represent the initial
volatile abundance of the giant-planet forming region of the solar
nebula, the resulting \acetylene/HCN, \acetylene/CO, and HCN/CO ratios
agree to within a factor of two with those found for GV Tau S, IRS 46,
massive YSOs, and several of the Markwick model results.

Also of interest is \methane, another particularly volatile molecule
for which the abundance (relative to water) has been observed to be
highly variable in comets (by over an order of magnitude).  While the
magnitude of the abundance variation is similar to that found for CO
in comets, the two molecules are not correlated \citep{gibb03}.  In
comets \methane/CO is found to vary from $\sim$0.05--0.8, much higher
than in HL Tau ($<$0.0002 in the gas phase) or GV Tau S ($<$0.0037),
and lower than the \methane/CO$\sim$1 found by \citet{marc02}.
\methane/HCN is found to vary from $\sim$2--5 among comets yet is
$<$0.6 toward GV Tau S.  The reasons for these compositional
differences are unknown and more work must be done to characterize
volatiles toward comets and low mass star forming regions.

\section{Conclusion}

We have measured column densities and rotational temperatures for CO,
HCN, and \acetylene\ and an upper limit for \methane\ toward GV Tau S.
We find that the absorptions are consistent with an origin in the
inner region of a protoplanetary disk.  This conclusion is further
strengthened by the lack of absorptions due to organic species toward
the infrared companion to GV Tau S, though gas phase CO is present.
The upper limit for HCN toward the IRC may suggest compositional
differences between the two objects, though a different temperature or
viewing geometry cannot be ruled out. We note that GV Tau is variable
in the near-infrared, likely due to inhomogeneities in the
circumbinary material or outer disk.  Future observations at M and L
band are planned to test whether the variability affects the column
densities and abundances of the gas phase species reported in this
paper.

We find that the abundances of HCN and \acetylene\ relative to each
other and to CO are similar to those found among the comet population,
that found toward low mass object IRS 46 by \citet{lahu06}, and
consistent with the disk model by \citet{marc02}.  \methane, on the
other hand, appears to be underabundant in the young stars sampled to
date when compared to comets in our own solar system.  This exciting
result illustrates the feasibility of detecting minor volatile
constituents toward low mass young stars from the ground, a study
which is vitally important to understanding how our own system
evolved.

\acknowledgements
  The data presented herein were obtained at the W.M. Keck
  Observatory, which is operated as a scientific partnership among the
  California Institute of Technology, the University of California and
  the National Aeronautics and Space Administration.  The Observatory
  was made possible by the generous financial support of the W.M. Keck
  Foundation. TWR was supported by NSF Astronomy grant AST02-05581 and
  AST05-07419.  SB was supported in part by a NASA Michelson
  Fellowship (2005-2006).  ELG was supported by NSF Astronomy grant
  AST-0507419.

\clearpage

\begin{table}
\begin{center}
\caption{Observing Log \label{table1}}
\begin{tabular}{lccc}
\tableline
Date & Instrument Setting & Wavelength Coverage & Integration Time  \\
     & & & (s) \\
\tableline
03/18/2003 & MW1 & 2118--2153~\wvn\ (order 16) & 240 \\
03/18/2003 & MW2 & 2094--2127~\wvn\ (order 16) & 240 \\
08/05/2003 & K1\tablenotemark{a}  & 4202--4267~\wvn\ (order 32) & 1200 \\
           &     & 4333--4397~\wvn\ (order 33) &      \\
08/05/2003 & K2\tablenotemark{a}  & 4265--4326~\wvn\ (order 33) & 1200 \\
02/17/2006 & KL1 & 3024--3072~\wvn\ (order 23) & 240 \\
           &     & 3286--3337~\wvn\ (order 25) &     \\
02/18/2006 & KL2\tablenotemark{a} & 2985--3028~\wvn\ (order 23) & 240 \\
           &     & 3243--3290~\wvn\ (order 25) &     \\
\tableline
\end{tabular}
\tablenotetext{a}{Note: The IRC was not in the slit in this observation.}
\end{center}
\end{table}

\clearpage

\begin{deluxetable}{lccc}
\tablecolumns{4}
\tabletypesize{\scriptsize}
\tablecaption{Column Densities and Rotational Temperatures for Molecules} 
\label{tbl2}
\tablewidth{0pt}
\tablehead{
\colhead{Molecule} & \colhead{Column Density} & \colhead{T$_{rot}$} & \colhead{Abundance Relative to $^{12}$CO} \\
& \colhead{(10$^{16}$ \pcmsq)} & \colhead{(K)} & \colhead{(\%)}}
\startdata
GV Tau S  &  & & \\
\hline
$^{12}$CO  & 590$\pm$120 & 200$\pm$40 & --- \\
$^{13}$CO  & 11          & 260$\pm$20 & 1.9  \\
C$^{18}$O  & 1.4$\pm$0.5 & 140$\pm$50 & 0.24  \\
HCN        & 3.7$\pm$0.3 & 115$^{+11}/_{-10}$ & 0.63 \\
\acetylene & 7.3$^{+0.1}/_{-0.2}$ & 170$^{+19}/_{-16}$ & 1.2 \\
\methane   & $<$2.2     &  --- & $<$0.37 \\
\hline\hline
Infrared Companion & & & \\
\hline
$^{12}$CO  & $\sim$300\tablenotemark{a}  & --- & --- \\
$^{13}$CO  & 5.5 & 220$\pm$40  & 54\tablenotemark{a} \\
HCN        & $<$0.48 & --- & $<$0.16 \\
\hline
\enddata
\tablenotetext{a}{Assuming $^{12}$CO/$^{13}$CO ratio is the same as for GV Tau S}
\end{deluxetable}

\clearpage

\begin{deluxetable}{lccccc}
\tablecolumns{6}
\tabletypesize{\scriptsize}
\tablecaption{HCN Line Positions and Equivalent Widths} 
\label{tbl2}
\tablewidth{0pt}
\tablehead{
\colhead{Line ID} & \colhead{$\nu_{rest}$} & \colhead{$\nu_{shift}$} &
\colhead{v$_{rad}$} & \colhead{\%T} & \colhead{W$\pm$dW} \\
& \colhead{(\wvn)} & \colhead{(\wvn)} & \colhead{(km s$^{-1}$)} & \colhead{(\wvn)}}
\startdata
P7 & 3290.35 & 3289.81 & 49 & 0.91 & 0.0154(0.0016) \\
P5 & 3296.49 & 3295.92 & 52 & 0.91 & 0.0184(0.0016) \\
P4 & 3299.53 & 3299.01 & 48 & 0.95 & 0.0183(0.0016) \\
P3 & 3302.53 & 3302.02 & 48 & 0.97 & 0.0162(0.0015) \\
P2 & 3305.54 & 3304.97 & 52 & 0.98 & 0.0136(0.0015) \\
R1 & 3317.33 & 3316.77 & 50 & 0.95 & 0.0128(0.0016) \\
R2 & 3320.22 & 3319.69 & 48 & 0.98 & 0.0179(0.0015) \\
R3 & 3323.09 & 3322.54 & 50 & 0.89 & 0.0212(0.0016) \\
R5 & 3328.78 & 3328.25 & 47 & 0.95 & 0.0222(0.0016) \\
R6 & 3331.59 & 3331.07 & 47 & 0.98 & 0.0236(0.0015) \\
\enddata

\end{deluxetable}

\clearpage

\begin{deluxetable}{lcccc}
\tablecolumns{5}
\tabletypesize{\scriptsize}
\tablecaption{\acetylene\ Line Positions and Equivalent 
Widths} \label{tbl3}
\tablewidth{0pt}
\tablehead{
\colhead{Line ID} & \colhead{$\nu_{rest}$} & 
\colhead{$\nu_{shift}$} & \colhead{\%T} 
& \colhead{W$\pm$dW} \\
& \colhead{(\wvn)} & \colhead{(\wvn)} & \colhead{(\wvn)}}
\startdata
$\nu_3$ band & & & &\\
R4  & 3306.48 & 3305.94 & 0.96 & 0.0067(0.0020) \\
P2  & 3290.13 & 3289.59 & 0.86 & 0.0058(0.0014) \\
P3  & 3287.75 & 3287.22 & 0.96 & 0.0044(0.0019) \\
P11 & 3268.48 & 3267.94 & 0.97 & 0.0077(0.0014) \\
$\nu_2+(\nu_4+\nu_5)^0_+$ band & & & &\\
R15 & 3318.35 & 3317.81 & 0.90 & 0.0037(0.0016) \\
R12 & 3311.71 & 3311.17 & 0.88 & 0.0041(0.0017) \\
R10 & 3307.22 & 3306.68 & 0.95 & 0.0049(0.0018) \\
R7  & 3300.42 & 3299.88 & 0.91 & 0.0084(0.0016) \\
R5  & 3295.84 & 3295.31 & 0.95 & 0.0124(0.0014) \\
R3  & 3291.23 & 3290.69 & 0.86 & 0.0047(0.0016) \\
P5  & 3270.05 & 3269.52 & 0.94 & 0.0044(0.0016) \\
P6  & 3267.66 & 3267.12 & 0.94 & 0.0105(0.0015) \\
P7  & 3265.26 & 3264.72 & 0.78 & 0.0088(0.0018) \\
P8  & 3262.84 & 3262.31 & 0.95 & 0.0059(0.0014) \\
P9  & 3260.43 & 3259.89 & 0.84 & 0.0077(0.0017) \\
P11 & 3255.56 & 3255.03 & 0.84 & 0.0046(0.0017) \\
P12 & 3253.12 & 3252.59 & 0.95 & 0.0025(0.0014) \\
P13 & 3250.66 & 3250.13 & 0.98 & 0.0053(0.0013) \\
P14 & 3248.20 & 3247.67 & 0.95 & 0.0026(0.0015) \\
P18 & 3251.09 & 3250.56 & 0.97 & $<$0.0014 \\
\enddata

\end{deluxetable}

\clearpage

\clearpage

\clearpage

\begin{figure}
\epsscale{1.0} 
\plotone{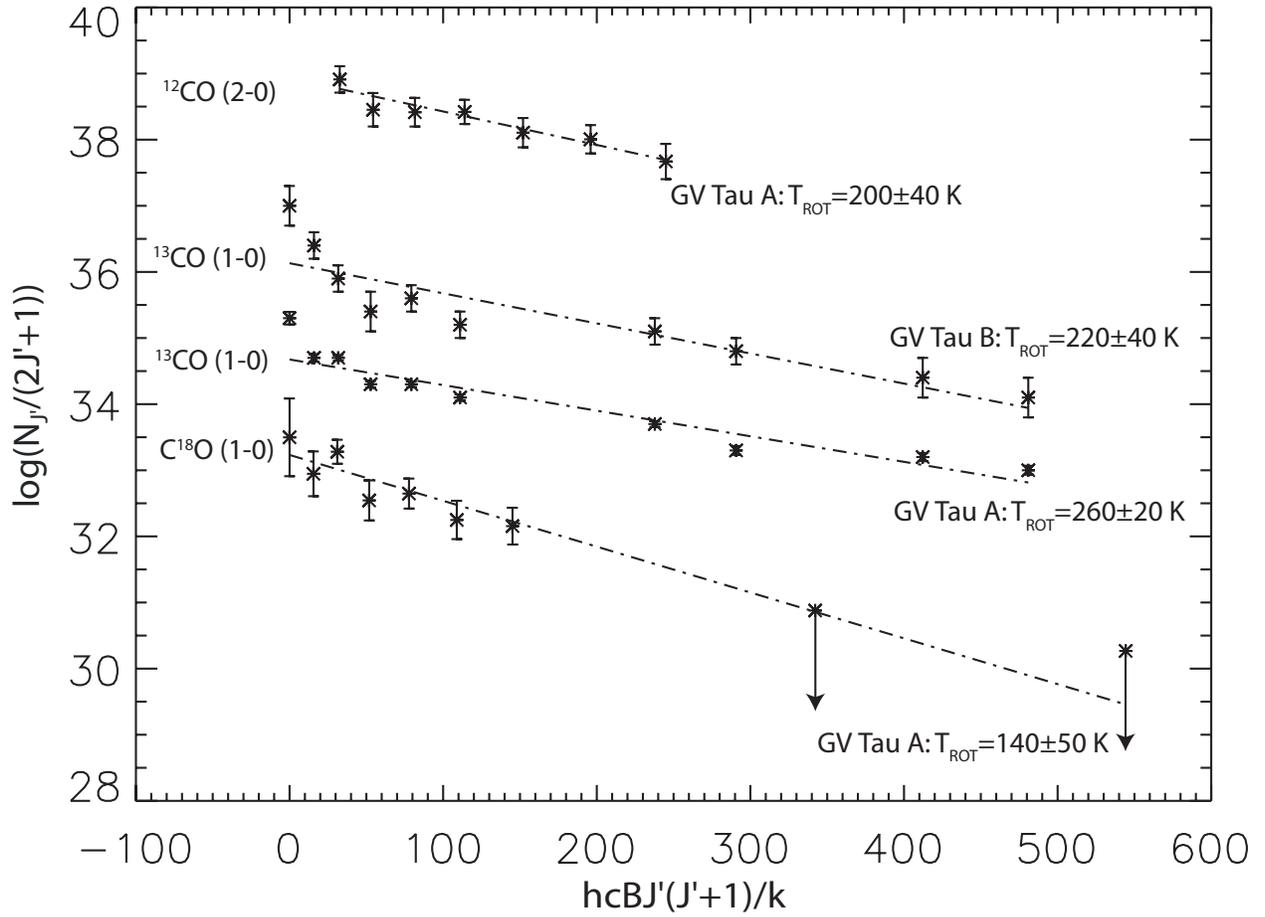}
\caption{Population diagram of the $^{12}$CO (2-0), 
$^{13}$CO (1-0) and C$^{18}$O (1-0) lines. \label{fig3}}
\end{figure}

\clearpage

\begin{figure}
\epsscale{1.0}
\plotone{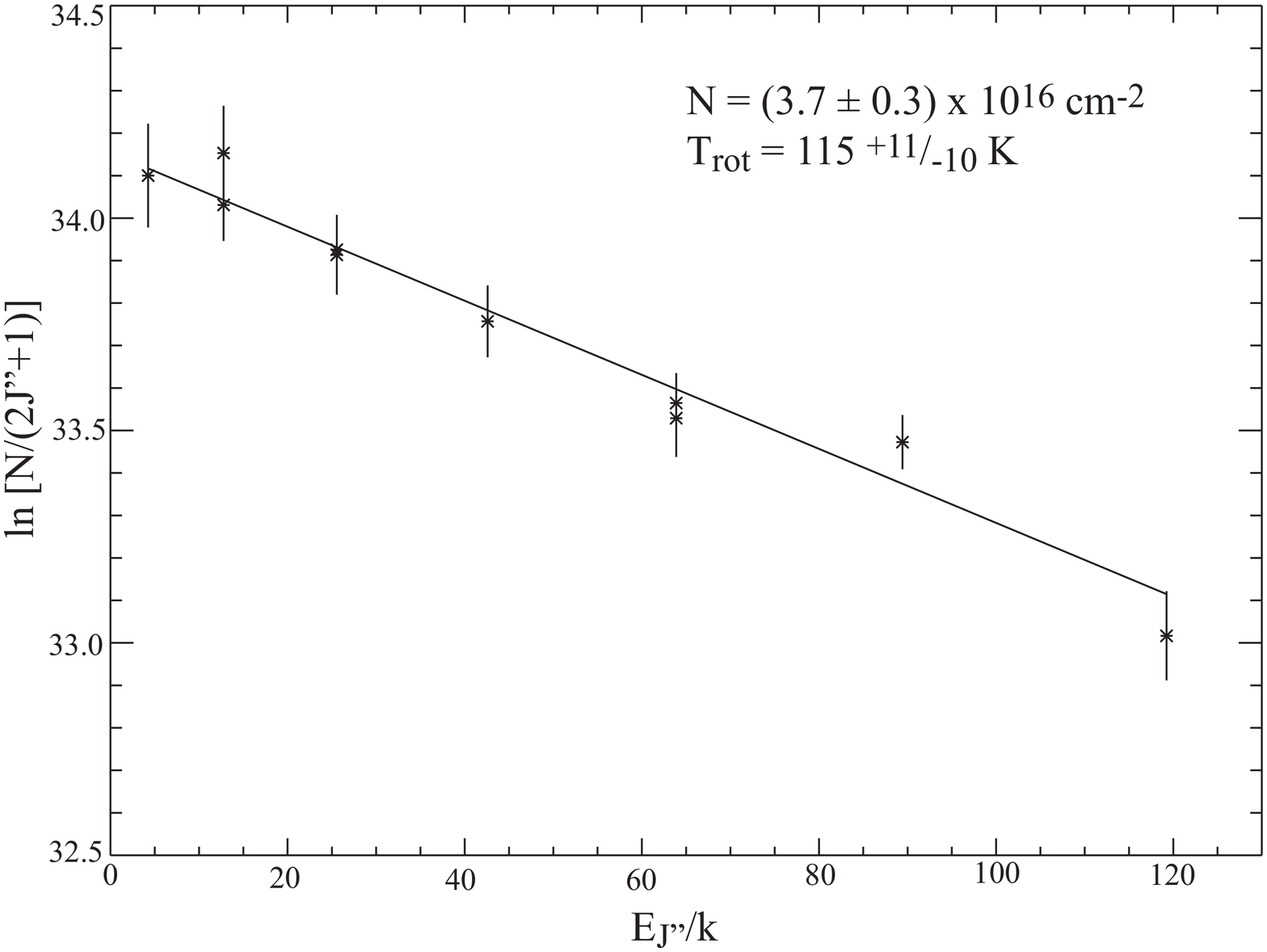}
\caption{Population diagram for HCN absorption toward GV Tau
S. \label{fig4}}
\end{figure}

\clearpage

\begin{figure} 
\epsscale{1.0} 
\plotone{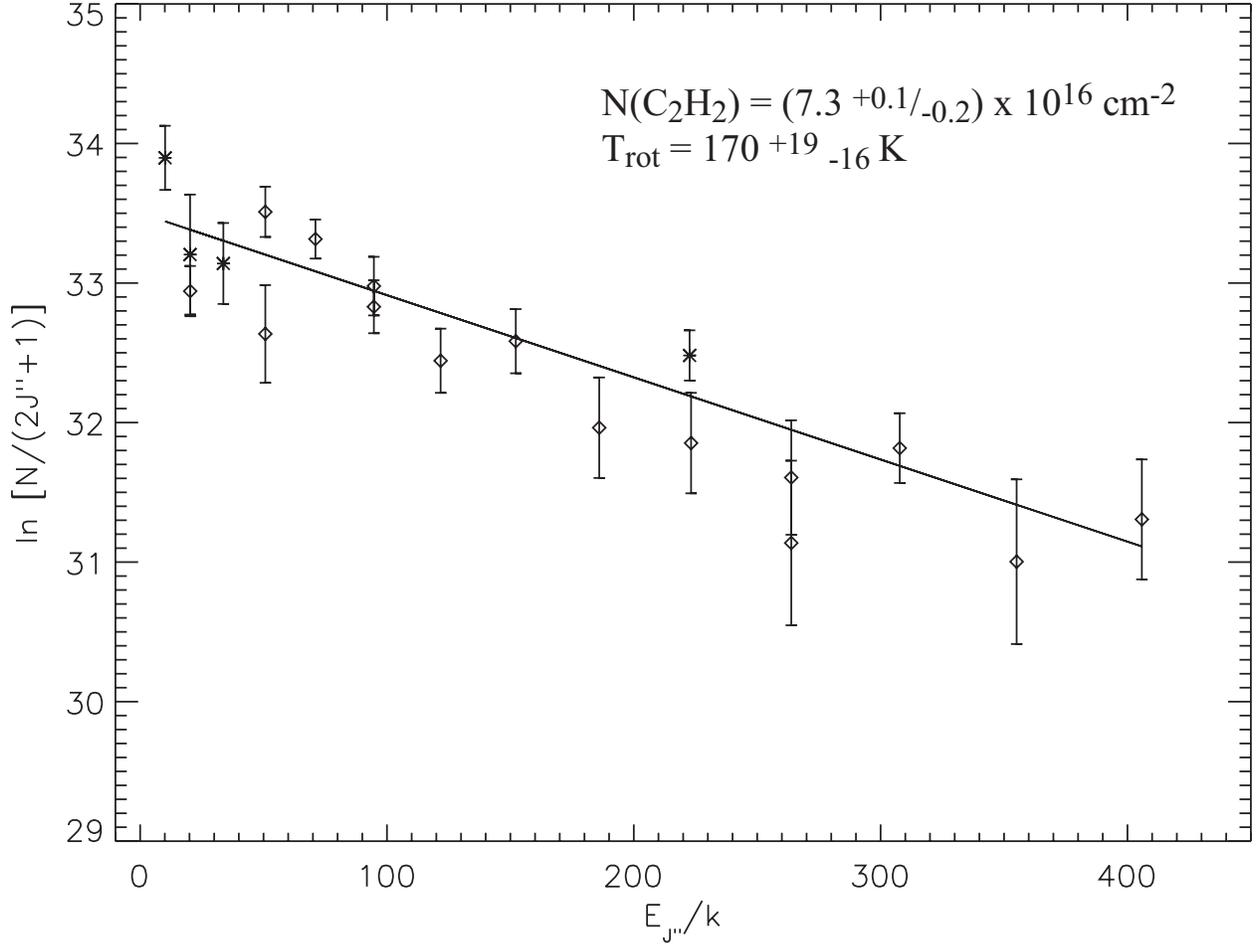} 
\caption{Population diagram for \acetylene\ absorption toward GV Tau
  S.  Asterisks are data points from the $\nu_3$ branch, diamonds are
  from the $\nu_2$ + ($\nu_4$ + $\nu_5$)$^0_+$ branch. \label{fig5}}
\end{figure}

\clearpage

\begin{figure} 
\epsscale{1.0} 
\plotone{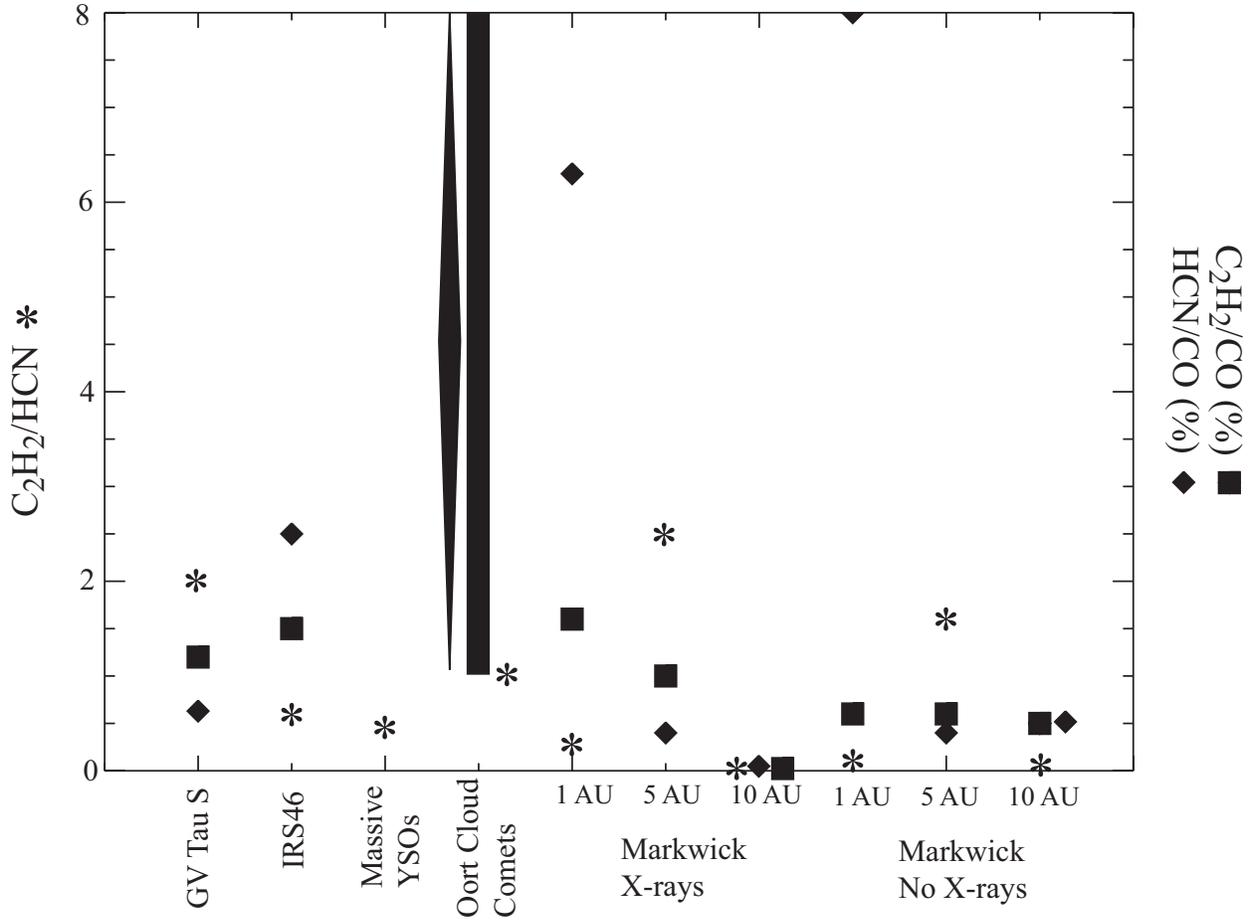} 
\caption{Figure comparing abundances toward GV Tau S, IRS 46, Massive YSOs, Oort Cloud Comets, and the chemical model of \citet{marc02}.  Asterisks indicate the \acetylene/HCN ratio.  Filled squares and diamonds indicate the abundances relative to CO of \acetylene\ and HCN, respectively. \label{fig6}} 
\end{figure}

\end{document}